\documentclass{mn2e}
\usepackage{psfig}

\def\AaA{A\&A} 
\def\AAR{A\&AR}

\def\ApJ{ApJ}

\def\ApJS{ApJS}

\def\ARAA{ARA\&A}

\def\JCP{J.~Comp.~Phys.}
\def\MN{MNRAS}
\def\Nat{Nature}

\def\RMP{Rev.~Mod.~Phys.}
\def\etal{{\rm et al.\thinspace}}

\def\TxLx{$T_{\rm ew}-L_{\rm bol}$\thinspace}
\def\LxTx{$L_{\rm bol}-T_{\rm ew}$\thinspace}
\def\eg{{\rm e.g.\ }}
\def\ie{{\rm i.e.\ }}
\def\Chandra{\textit{Chandra}\thinspace}

\def\Sone{\textit{E1}}
\def\Stwo{\textit{E2}}
\def\Sthree{\textit{E3}}
\def\Sfour{\textit{E4}}
\def\Uone{\textit{U1}}
\def\Utwo{\textit{U2}}
\def\Uthree{\textit{U3}}
\def\Cone{\textit{C1}}
\def\Ctwo{\textit{C2}}
\def\Cthree{\textit{C3}}
\def\Rone{\textit{R1}}
\def\Rtwo{\textit{R2}}

\begin{document}

\title[Cluster Mergers]
{Hydrodynamic simulations of merging clusters of galaxies}
\author[B. W. Ritchie and P. A. Thomas]
{Benedict W. Ritchie$^{1,2}$\thanks{Email: Ben\_Ritchie@uk.ibm.com}
and Peter A. Thomas$^{1}$ \\
$^{1}$ Astronomy Centre, School of Chemistry, Physics and Environmental Science, 
University of Sussex, Falmer, Brighton BN1 9QJ \\
$^{2}$ IBM United Kingdom Laboratories, Hursley Park, Winchester, Hampshire SO21 2JN \\}

\date{Accepted ---. Received ---; in original form ---}

\maketitle

\begin{abstract}

We present the results of high-resolution AP$^3$M+SPH simulations of
merging clusters of galaxies. We find that the compression and
shocking of the core gas during a merger can lead to large increases
in bolometric X-ray luminosities and emission-weighted temperatures of
clusters. Cooling flows are completely disrupted during equal-mass
mergers, with the mass deposition rate dropping to zero as the cores
of the clusters collide. The large increase in the cooling time of the
core gas strongly suggests that cooling flows will not recover from
such a merger within a Hubble time. Mergers with subclumps having
1/8$^{\rm th}$ of the mass of the main cluster are also found to
disrupt a cooling flow if the merger is head-on.  However, in this
case the entropy injected into the core gas is rapidly radiated away
and the cooling flow restarts within a few Gyr of the merger.  Mergers
in which the subcluster has an impact parameter of 500kpc do not
disrupt the cooling flow, although the mass deposition rate is reduced
by $\sim 30\%$.  Finally, we find that equal mass, off-centre mergers
can effectively mix gas in the cores of clusters, while head on
mergers lead to very little mixing. Gas stripped from the outer layers
of subclumps results in parts of the outer layers of the main cluster
being well mixed, although they have little effect on the gas in the
core of the cluster. None of the mergers examined here resulted in the
ICM being well mixed globally.

\end{abstract}

\begin{keywords}
methods: numerical - hydrodynamics - cooling flows - X-rays : galaxies : clusters
\end{keywords}

\section{Introduction}

In the Cold Dark Matter scenario, clusters of galaxies form through a
succession of mergers of smaller subclusters and groups. Major mergers
can be the most energetic events in the Universe since the Big Bang,
with hydrodynamic shocks dissipating much of the kinetic energy of the
collision ($10^{63-64}$ ergs) into the Intracluster Medium (ICM). The
theoretical picture is supported by X-ray observations, which suggest
that many clusters of galaxies are either undergoing mergers or have
experienced merger events in the recent past. \textit{ROSAT}
observations generally focussed on distortion of the X-ray isophotes
(\eg White, Briel \& Henry 1993) and the presence of X-ray
substructure (Forman \& Jones 1994; Buote \& Tsai 1996), while
evidence for shocks in the ICM was seen in temperature maps created
from \textit{ASCA} observations (\eg Arnaud \etal 1994). The launch of
\Chandra and \textit{XMM-Newton} has provided a wealth of
observational data, including high-resolution temperature maps of
shocks in merging clusters (Markevich \& Vikhlinin 2001) and the
surprising discovery of sharp surface brightness discontinuities
between the hot cluster gas and moving ``cold fronts'' (Vikhlinin,
Markevitch \& Murray 2001), which are thought to be the cores of
subclusters which have survived the merger process (Markevitch \etal
2000).

Modern cosmological simulations (Bertschinger 1998 and references
therein; see also Frenk \etal 1999) are capable of following the
evolution of large volumes of the universe from early times until the
present day, but are not particularly suited to modeling mergers as
the resolution of individual clusters is limited unless techniques
such as resimulation (\eg Eke, Navarro \& Frenk 1998) are
used. Cosmological simulations are also hard to analyse as the cluster
does not necessarily have time to return to hydrostatic equilibrium
between mergers, making it hard to separate one merger from the
next. An alternative is to study the merger of idealized systems,
which, while less realistic than the full cosmological approach, have
a number of advantages. Firstly, the objects participating in the
merger are fully formed, and so there is no need to simulate the large
volumes of space necessary to capture the evolution of a cluster of
galaxies from high redshift. We can therefore use a smaller volume in
which the merging systems are well resolved, and this also has
implications for our hydrodynamic scheme; this is discussed in
Section~\ref{sec:method}. As we are studying a single merger event
between two objects which are initially in hydrostatic equilibrium
changes in the final object are easily quantified, and the controlled
initial conditions allow parameter space to be explored. While full
cosmological simulations are ultimately more realistic, the simplified
picture represents the best chance of understanding the physics
underlying the merger process.

Several authors have presented results from such simulations.  Head-on
mergers between King-model clusters of unequal mass containing gas and
dark matter were studied by Roettiger, Burns \& Loken (1993) and
Roettiger, Loken \& Burns (1997, hereafter RLB97) using a
finite-difference code to model the hydrodynamics and an $N$-body code
to model the evolution of the gravitational potential, while Schindler
\& M\"{u}ller (1993) used a similar approach, although with less
controlled initial conditions. However, these simulations treated the
baryonic component as being massless, limiting the results to very
gas-poor clusters. Pearce, Thomas \& Couchman (1994, hereafter PTC94)
examined head-on mergers of systems containing gas and dark matter
using an adaptive particle-particle, particle-mesh (AP$^3$M; Couchman
1991) code with the gas being modeled using Smoothed-Particle
Hydrodynamics (SPH; Monaghan 1992), with their results focusing mainly
on the evolution of the core of the cluster rather than the observable
properties of the merger.

More recently Roettiger, Stone \& Mushotzky (1998) have used a
hydrodynamic code based on the Piecewise-Parabolic Method (PPM;
Collela \& Woodward 1984) coupled to a Particle-Mesh $N$-body code to
examine the evolution of the cluster A754, suggesting that its
unusual X-ray morphology (\eg Henry \& Briel 1995) is the result of a
recent ($<0.5$Gyr), slightly off-centre merger. Ricker (1998) also
used PPM to simulate off-centre cluster mergers, although these
simulations do not include the dark matter which typically dominates
the dynamics of clusters and can make a significant difference to the
end state of the merger process (PTC94; see also
Section~\ref{sec:struct}). Finally, Takizawa (1999; 2000) carried out
simulations of merging clusters using SPH, with gravitational forces
calculated using a tree algorithm (Barnes \& Hut 1986). The
simulations of PTC94 and Takizawa (1999) are the most similar to our
own and provide a useful comparison, although both were carried out at
lower resolution than the results described here.

In this paper we present results from high-resolution (131072
particle) simulations of the merger of idealized clusters of galaxies
containing both dark matter and gas, using an AP$^3$M+SPH
code. Off-centre mergers and mergers between unequal-mass systems are
also examined, with reference to the observable properties of clusters
of galaxies. The layout of the paper is as follows; in
Section~\ref{sec:method} we describe our hydrodynamic scheme and the
generation of initial conditions, while results are described in
Section~\ref{sec:results}. The significance of the results are
discussed in Section~\ref{sec:discuss}, and conclusions are drawn in
Section~\ref{sec:conc}.

\section{Method}
\label{sec:method}

%Bloody LaTeX seems to need this to be here in order for it to
%appear somewhere sensible in the text.

\begin{table*}
\begin{minipage}{17.6cm}
\begin{center}
\begin{tabular}{|c|c|c|c|c|c|c|c|c|c|}
\hline
Run no. & $N_p$ & ${\rm M}_{\rm dm} ({\rm M}_{\odot})$ & ${\rm M}_{\rm gas}
({\rm M}x_{\odot})$& $\rho_0 ({\rm cm}^{-3})$& $M_1/M_2$ & $b/r_{\rm core}$ &
$\Delta L_{\rm x}/L_{\rm x}$ & $\Delta T_{\rm ew}/T_{\rm ew}$ & $\Delta E/E$ \\
\hline
\Sone & 131072 & $1.25 \times 10^{10}$ & $1.13\times10^9$ & $10^{-3}$ &
1. & 0.  & 7.99 & 3.31 & $5\times10^{-4}$ \\
\Stwo & 131072 & $1.25 \times 10^{10}$ & $1.13\times10^9$& $10^{-3}$&
1. & 2.5 & 7.01 & 3.17 & $6\times10^{-4}$ \\
\Sthree & 131072 & $1.25 \times 10^{10}$ & $1.13\times10^9$& $10^{-3}$&
1. & 5.  & 5.01 & 2.80 & $1\times10^{-4}$ \\
\Sfour & 131072 & $1.25 \times 10^{10}$ & $1.13\times10^9$& $10^{-3}$&
1. & 10. & 2.59 & 1.86 & $3\times10^{-4}$ \\
\hline
\Uone & 73728  & $1.25 \times 10^{10}$& $1.13\times10^9$& $10^{-3}$&
8. & 0.  & 2.26 & 1.71 & $8\times10^{-4}$ \\
\Utwo & 73728  & $1.25 \times 10^{10}$& $1.13\times10^9$& $10^{-3}$&
8. & 2.5 & 1.95 & 1.55 & $9\times10^{-4}$\\
\Uthree & 73728  & $1.25 \times 10^{10}$& $1.13\times10^9$& $10^{-3}$&
8. & 5.  & 1.52 & 1.32 & $1.1\times10^{-3}$\\
\hline
\Cone & 131072 & $1.25 \times 10^{10}$& $1.13\times10^9$& $10^{-2}$&
1. & 0.  & 4.94 & 4.34 & $8\times10^{-4}$\\
\Ctwo & 73728  & $1.25 \times 10^{10}$& $1.13\times10^9$& $10^{-2}$&
8. & 0.  & 2.18 & 1.83 & $4.4\times10^{-3}$\\
\Cthree & 73728  & $1.25 \times 10^{10}$& $1.13\times10^9$& $10^{-2}$&
8. & 2.5 & 1.09 & 1.08 & $3.9\times10^{-3}$\\
\hline 
\Rone & 32768  & $5 \times 10^{10}$& $4.5\times10^9$& $10^{-3}$& 1. &
0.  & 5.01 & 3.34 & $2.0\times10^{-3}$\\ 
\Rtwo & 8192   & $2 \times 10^{11}$& $1.8\times10^{10}$& $10^{-3}$&
1. & 0.  & 3.60 & 2.94 & $2.6\times10^{-3}$\\
\hline
\end{tabular}
\caption{Details of the simulations examined in
Section~\ref{sec:results}. Listed are the run number, the total number
of particles in the simulation (Dark matter + Gas), the mass of each
dark matter particle, the mass of each gas particle, the core
density, the ratio of masses of the two clusters $M_1/M_2$, the impact
parameter $b$, the ratio of the peak bolometric X-ray luminosity to
the bolometric X-ray luminosity 6Gyr before the
merger $\Delta L_{\rm x}/L_{\rm x} = L_{\rm t=0}/L_{t=-6Gyr}$, the
ratio of the peak emission-weighted temperature to the
emission-weighted temperature 6Gyr times before the merger
$\Delta T_{\rm ew}/T_{\rm ew} = T_{\rm t=0}/L_{\rm t=-6Gyr}$ and the
energy conservation during the simulation $\Delta E/E$.}
\label{tab:simulations}
\end{center}
\end{minipage}
\end{table*}

The simulations described here have been carried out using
HYDRA\footnote{This code is in the public domain and can be downloaded
from http://hydra.sussex.ac.uk/ or http://hydra.mcmaster.ca/.}
(Couchman, Thomas \& Pearce 1995), an $N$-body, hydrodynamics code
which combines an AP$^3$M $N$-body algorithm (Couchman 1991) with
SPH. Tests of our code can be found in Couchman \etal
(1995), Thacker \etal (2000) and Ritchie \& Thomas (2001).

In SPH, the equations of motion for a compressible fluid (\eg Landau
\& Lifshitz 1959) are solved using a Lagrangian formulation in which
the fluid is partitioned into elements, a subset of which are
represented by particles of known mass $m$ and specific energy
$\epsilon$\footnote{The specific energy $\epsilon$ is related to the
gas temperature $T$ by $\epsilon = 3k_{\rm B}T/2\mu m_{\rm H}$, where
$k_{\rm B}$ is Boltzmann's constant, $m_{\rm H}$ is the mass of the
hydrogen atom and $\mu = 0.6$ is the relative molecular
mass.}. Continuous fields are represented by interpolating between
particles using a smoothing kernel, which is normally defined in terms
of a sphere containing a fixed number of neighbours, centred on the
particle in question.  The radius of the smoothing sphere is adjusted
so as to keep the neighbour count approximately constant, making SPH
adaptive in both space and time. The particle nature of SPH means that
there is no grid to constrain the geometry or dynamic range of the
system being studied, and allows SPH to be easily integrated with many
$N$-body solvers. However, unlike PPM, SPH requires an artificial
viscosity to convert relative motion to heat; we use a pairwise
artificial viscosity (Monaghan \& Gingold 1983) as described by
Thacker \etal (2000).

SPH is often criticised for capturing shocks poorly in comparison with
modern high-order Godunov-type schemes, and for having poor resolution
in low-density regions, which are represented by relatively few
particles. While both of these criticisms can be true, they are of
limited importance for the simulations considered here. Steinmetz \&
M\"{u}ller (1993) find that the shock capturing ability of SPH is
closely linked to the number of particles in the system being modeled,
with SPH giving accurate results if in excess of $10^4$ particles are
used to model three-dimensional problems. This figure is out of reach
of most cosmological simulations, but is easily achievable with our
simplified approach, and the initial conditions are generated with the
constraints of Steinmetz \& M\"{u}ller (1993) in mind. We will not be
able to follow the propagation of shocks into the lower-density outer
regions of the cluster as well as finite-difference methods, but the
X-ray luminosity of such regions will be low and so the limitations of
our hydrodynamical method will not impact any potentially observable
properties of the system.

Radiative cooling is implemented in our code by adding a sink term
$\xi$ to the Energy equation, where $\xi$ is the emissivity (the
emission rate per unit volume), interpolated from the cooling function
of Sutherland \& Dopita (1993). Cooling during shock-heating (see
Hutchings \& Thomas 2000) is minimized by applying the artificial
viscosity prior to the gas being allowed to cool.  The cooling is
assumed to occur at constant density, with the time-step ensuring that
this condition is approximately satisfied, as described in Thomas \&
Couchman (1992).  Particles which have cooled to 10$^4$K, the
temperature at which the cooling function drops to zero, are converted
to collisionless `star' particles to avoid a build-up of cold gas that
can cause the SPH algorithm problems (Pearce \etal 2001; Ritchie \&
Thomas 2001). We neglect the effect of thermal conductivity, which is
known to be strongly suppressed in clusters of galaxies (\eg Fabian,
Nulsen \& Canizares 1991; Vikhlinin \etal 2001).

\subsection{Initial Conditions}
\label{sec:ic}

Our idealized clusters initially have gas and dark matter density
profiles given by a Hubble profile
\begin{equation}
\rho(r) = \frac{\rho_0}{\left [1+(r/r_c)^2 \right]^{3/2}},
\label{eq:hubprof}
\end{equation}
where $\rho_0$ is the central mass density and $r_c$ is the cluster
core radius. This is similar to the gas density profile deduced from
the X-ray surface brightness of clusters of galaxies, and has been
widely used by other authors simulating cluster mergers (\eg PTC94;
RLB97; Ricker 1998; Takizawa 1999). $N$-body simulations (\eg Moore
\etal 1999) suggest that the dark matter is more centrally
concentrated than that given by Equation~\ref{eq:hubprof}, and Ricker
\& Sarazin (2001) have recently carried out simulations in which the
dark matter density follows the Navarro, Frenk \& White (1997) profile
\begin{equation}
\rho(r) = \frac{\rho_s}{(r/r_s)(1+r/r_s)^2},
\end{equation}
where $\rho_s$ and $r_s$ are a scaling density and radius that are
dependent on the halo mass.  They find that in their simulations the
morphological changes, temperature jumps and gas velocities that are
similar to previous work in which the dark matter density is given by
Equation~\ref{eq:hubprof}, although the peak X-ray luminosity is
higher.

In our simulations, we truncate the density profile at $R=16r_c$, a
numerical compromise which allows the core to be well resolved yet be
located well within the outer boundary of the system. To set up the
density profile, we first place gas particles randomly within a
cubical simulation volume, which is then evolved at constant
temperature until spurious fluctuations arising from the initial
particle distribution have died away. Particles are then ordered in
terms of their distance from the centre of the box and are translated
radially to match the desired mass profile, ensuring that the gas
starts off close to a relaxed state. Finally, collisionless dark
matter particles with 9 times the gas mass are placed on top of each
gas particle.

Once particles have been placed, particle velocities must be set. Dark
matter velocities are drawn from a 3--D Gaussian distribution of width
$\sigma$ in each direction, where $\sigma$ is determined by solving the 
Jeans equation
\begin{equation}
\frac{1}{\rho} \frac{{\rm d}(\rho \sigma^2)}{{\rm d}r} = - \frac{GM}{r^2}.
\label{eq:jeans}
\end{equation}
Gas particles have their initial velocities set to zero and are given
a temperature equivalent to $\sigma^2$, so that
\begin{equation} 
\beta = \frac{\sigma^2}{k_{\rm B}T / \mu m_{\rm H}} = 1,
\end{equation} 
with particle densities set from Equation~\ref{eq:hubprof}. The sound
crossing time for the cluster is
\begin{equation}
t_{\rm sc} \equiv \frac{R}{c_s} = R \sqrt{\frac{\mu m_{\rm H}}{\gamma k_{\rm B} T_{\rm c}}} = 1.75{\rm Gyr.}
\end{equation}

\begin{figure}
$$\vbox{ \psfig{file=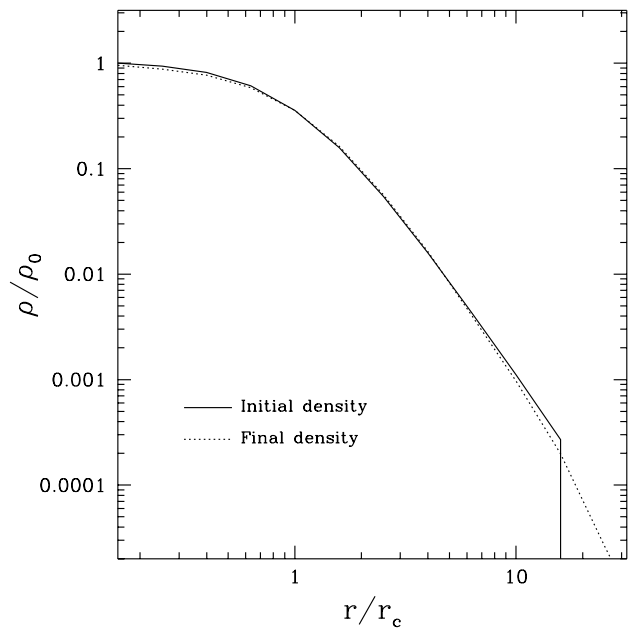,width=8.7cm} }$$
\caption{Radial density profiles in a single-cluster test run. The
initial density profile is plotted using a solid line, while the
density profile after 20 sound crossing times is plotted using a
dotted line. Both profiles have been normalised to $\rho_0$, the
initial core density.}
\label{fig:dpsingle}
\end{figure}

Figure~\ref{fig:dpsingle} shows the evolution of the density profile
over a period of 20 sound crossing times for an isolated system set up
with the density profile specified by Equation~\ref{eq:hubprof}. Some
expansion in the outer region due to the initially-truncated density
profile is visible, but these fluctuations are small and settle down
within a couple of sound crossings. Our initial conditions for
simulations of equal mass mergers simply take two such spheres,
displaced by a distance equal to the impact parameter $b$ in the
$y$-direction and a distance $2R$ in the $x$-direction. The initial
relative velocity in the $x$-direction is equal to the circular speed
\begin{equation}
\label{eq:vcoll}
v_{\rm coll} = \sqrt{\frac{GM}{R}}, 
\end{equation}
at the edge of the cluster, where $M$ is the mass of each cluster and
$R=16r_c$ as above. This corresponds to a parabolic orbit but, once
energy is dissipated in the encounter, leads to a bound system. The
effect of varying the encounter velocity is discussed in PTC94.

Mergers between systems of different mass are set up by using fewer
particles in the lower-mass cluster (\ie the mass per particle is kept
constant) and scaling the second cluster to match the desired radius
and velocity dispersion profile. We use the same impact parameter and
relative velocity as in the equal mass encounters.

\subsection{Simulations}

Details of the simulations are listed in Table~\ref{tab:simulations}.
With the exception of the simulations carried out in
Section~\ref{sec:res}, which examine the effect numerical resolution
has on our results, all simulations of equal-mass mergers presented
here use 32768 particles each of gas and dark matter in each cluster
(\ie an equal-mass merger simulation contains a total of 131072
particles). In simulations of unequal-mass mergers the particle mass
is kept constant, so that the subcluster contains $1/N$ fewer
particles, where $N = M_1/M_2$ is the ratio of cluster masses. We are
not attempting to simulate the evolution of clusters in a cosmological
context, and therefore our simulations use a non-expanding volume with
vacuum boundary conditions. Particles leaving the simulation volume
are removed, although the size of the box is such that very few
particles escape by the end of the simulation.  We use a gravitational
softening $r_s = 0.2r_c$, which is above the particle separation in
the core. This choice of softening is discussed in detail in Pearce,
Thomas \& Couchman (1993).

We use a gas particle mass of $1.25\times10^9 {\rm M}_{\odot}$ and a
dark matter mass $1.13\times10^{10} {\rm M}_{\odot}$, giving a total
cluster mass of $4\times 10^{14}{\rm M}_{\odot}$ and virial
temperature $\sim3$keV. The subclusters used in simulating unequal
mass mergers have a mass of $5\times10^{13}{\rm M}_{\odot}$ and virial
temperature $\sim0.9$keV. We take the core radius $r_c$ to be 100kpc,
giving a central gas density $\rho_0\sim10^{-3}$cm$^{-3}$, similar to
that used by RLB97. The bremsstrahlung cooling time
\begin{equation}
t_{\rm cool} = \frac{3}{2}\frac{k_{\rm B}T}{\Lambda(T)} \approx
8.5\times10^{10} \left(\frac{T}{10^8{\rm K}} \right)^{1/2} \left(
\frac{n}{10^{-3}{\rm cm}^{-3}} \right )^{-1} {\rm yr}
\label{eq:tcool}
\end{equation}
(Sarazin 1986), where $\Lambda(T) = \Lambda_0(T/K)^{1/2}$ and
$\Lambda_0 = 5.2\times10^{-28}$ erg cm$^3$ s$^{-1}$, is therefore in
excess of a Hubble time in the core of the cluster. While
this might suggest that radiative cooling will have little impact on
the results of our simulations, cooling times can drop below a Hubble
time when the density of the core gas increases as the cores collide
and radiative cooling makes a significant (20---40\%) difference to
our results; this is discussed further in Section~\ref{sec:lxtx}.  In
addition, we perform some simulations with a higher core density
$\rho_0 \sim 10^{-2}$cm$^{-3}$ (see Section~\ref{sec:cflow}) in which
the cooling time is much less than a Hubble time, although $t_{\rm
cool}$ remains greater than the sound crossing time $t_{\rm sc}$ in
all of our simulations. We therefore apply radiative cooling to all
the simulations performed here.

All of the simulations described here were carried out on a 700MHz
Intel Pentium-3 workstation. The simulations of mergers between equal
mass, low core density clusters take around 1800 timesteps and 15
hours of CPU time, while simulations of mergers between high core
density clusters take around 2500 timesteps and 21 hours of CPU time.

\section{Results}
\label{sec:results}

\subsection{Morphology}
\label{sec:morph}

\begin{figure*}
\begin{minipage}{17.6cm}
 \begin{center} \psfig{file=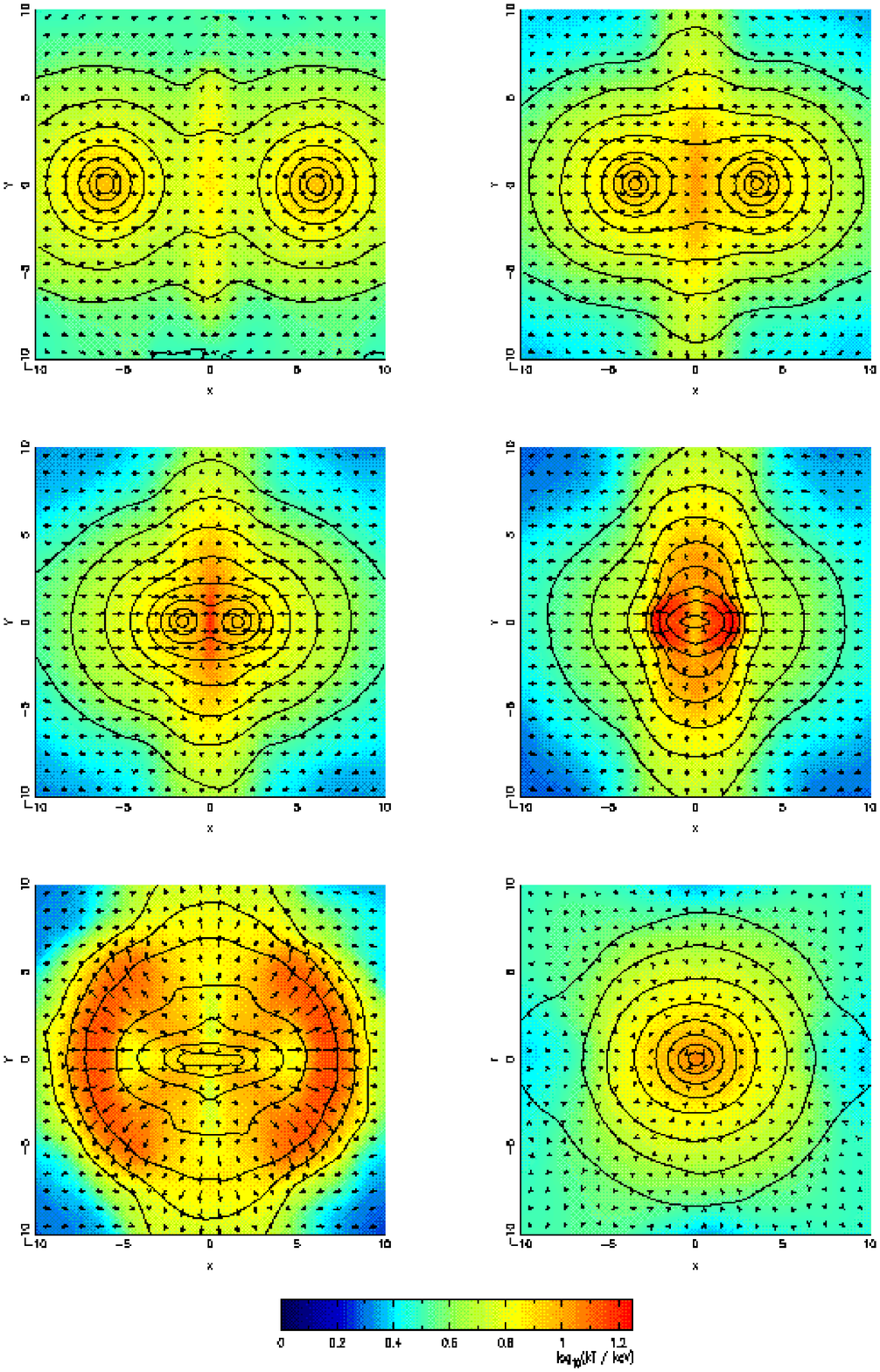,width=17.4cm}
   \caption{Snapshots of the head-on merger of two equal mass systems
   (run \Sone). Colours represent the logarithm of the gas temperature,
   contours represent logarithmic X-ray isophotes and arrows represent
   the velocity field. Panels are displayed at $t\simeq 0,3.5,6,7,8,10.$ and
   $20$Gyr.}
   \label{fig:emho} \end{center}
\end{minipage}
\end{figure*}

Figure~\ref{fig:emho} shows a sequence of snapshots of the
emission-weighted gas temperature (colours), X-ray surface brightness
(contours) and velocity fields (arrows) during the head on merger
between two systems with a masses of $4\times10^{14}{\rm M}_{\odot}$ (run
\Sone). The merging sequence is similar to that seen in other
simulations (\eg Ricker 1998; Takizawa 2000), and is therefore only
described briefly here.

As the outer layers of the two clusters start to interact a weak shock
forms (panels 1 \& 2), and gas is driven outwards in the plane
perpendicular to the collision axis with the gas velocity increasing
as it accelerates down the pressure gradient. As the cluster cores
approach this shock strengthens (panel 3), as does the outflow of gas
in the plane of the collision. The increase in gas temperature between
the two cores has become clearly visible in temperature maps of the
system, but as the bulk of the X-ray emission is coming from the cores
of the two clusters the shock still makes only a minor contribution to
the integrated emission-weighted temperature of the system (see
Section~\ref{sec:lxtx}). As the cluster cores collide (just before
panel 4), a strong arc-shaped shock is driven into the outer layers of
the cluster (visible in panel 4, and more clearly in panel
5). Meanwhile, gas in the core of the merged cluster goes through a
period of expansion driven by the dark matter, cooling adiabatically
to slightly less than the pre-collision temperature with the X-ray
isophotes forming a bar shape along the collision axis, as noted by
RLB97.  Finally, the dark matter turns around and recollapses to form
a spherically symmetric final object (panel 6).  Material ejected in
the earlier stages of the merger can be seen falling back towards the
core of the cluster.

\begin{figure*}
\begin{minipage}{17.6cm}
 \begin{center} \psfig{file=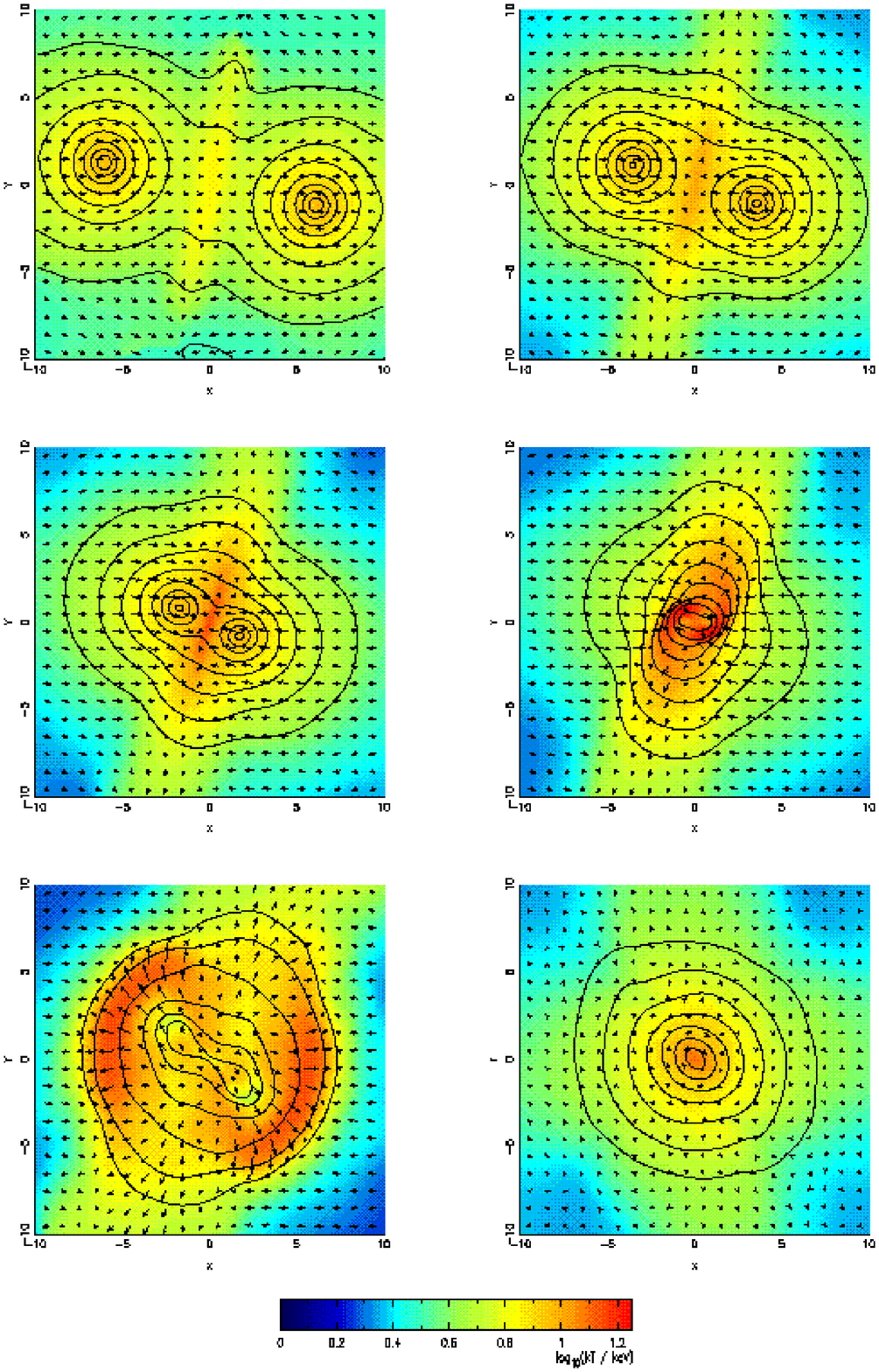,width=17.4cm}
   \caption{Snapshots of an off-centre merger with impact parameter
   $b=5$, between two equal mass systems (run \Sthree). Colours, contours
   and arrows are as in Figure~\ref{fig:emho}, and the same output
   times are used.}  \label{fig:emb5} \end{center}
\end{minipage}
\end{figure*}

\begin{figure}
$$\vbox{ \psfig{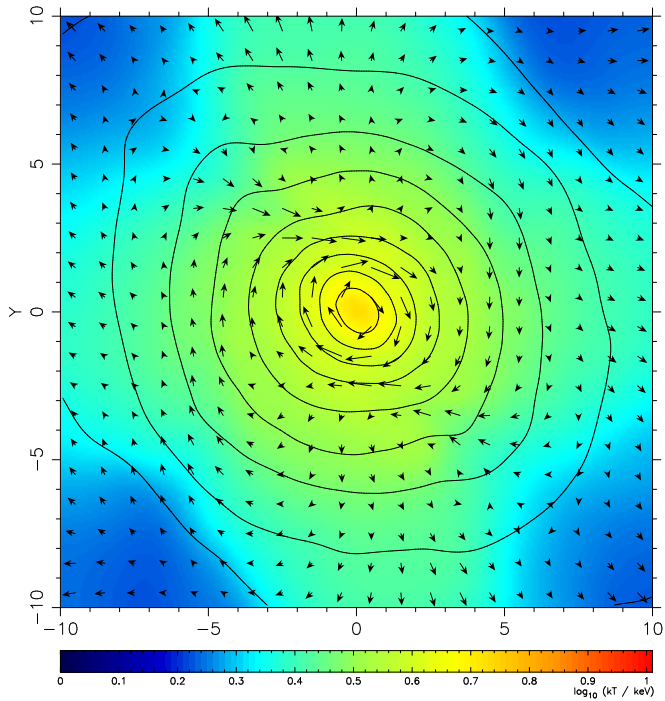} }$$
\caption{The final state of run \Sthree, the $b=5$ off-centre merger
shown in Figure~\ref{fig:emb5}. Greyscales, contours and arrows
represent the same quantities as in Figures~\ref{fig:emho}
and~\ref{fig:emb5}, but have been rescaled to better show the rotation
of the merger remnant.}
\label{fig:rotfinal}
\end{figure}

An off-centre merger with impact parameter $b=5$ (run \Sthree) is shown
in Figure~\ref{fig:emb5}. As before, the clusters have equal masses,
and the behaviour is similar to that seen in Figure~\ref{fig:emho}.
The shock that forms as the outer layers of the clusters interact is
now oblique, but the shock and outflow is otherwise similar to the
head-on merger. The picture changes somewhat as the cluster cores
interact, with the merger shock generated as the cores collide now
propagating with a spiral pattern, and the cores partially surviving
their first approach (clearly visible in panel 5), completing most of
an orbit before recollapsing and merging completely. The final state
of the cluster looks similar to that in the head-on case, although
there is now significant rotation of the core, shown in more
detail in Figure~\ref{fig:rotfinal}.

\subsection{The structure of the merger remnant}
\label{sec:struct}

\begin{figure}
$$\vbox{ \psfig{file=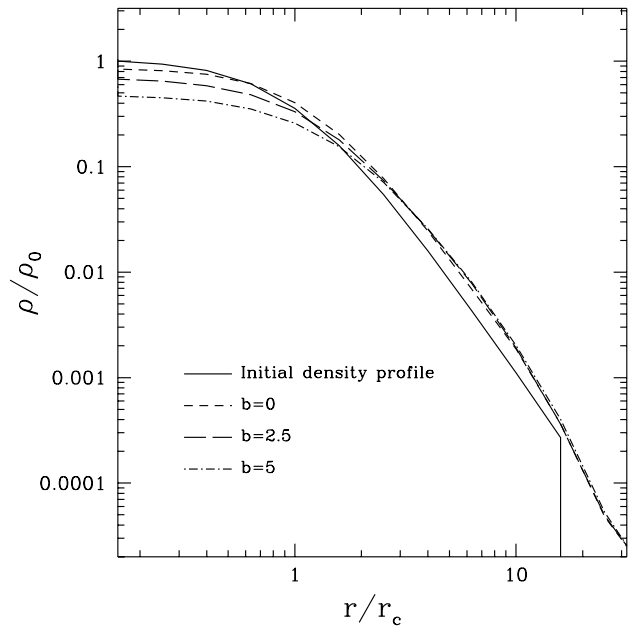,width=8.7cm} }$$
\caption{Radial gas density profiles of the merger remnant 10 sound
crossing times after the merger, for three equal mass mergers with
impact parameter $b=$0 (run \Sone, short-dashed line), 2.5 (run \Stwo,
long-dashed line) and 5 (run \Sthree, dot-dashed line). The initial
density profile (solid line) is plotted for reference. All profiles
have been normalised to $\rho_0$, the initial core density. Best fit
values for $\rho_0$ and $r_c$ are listed in Table~\ref{tab:hubfit}.}
\label{fig:rhoba}
\end{figure} 

\begin{table}
\begin{center}
\begin{tabular}{|c|c|c|c|}
\hline
Run no. & $b_{\rm impact}$ & $r_{\rm core}$ (kpc)& $\rho_{c} / \rho_0$ \\
\hline
Initial &  -  & 108 & 1  \\
\Sone    & 0   & 125 & 0.85 \\
\Stwo    & 2.5 & 135 & 0.65 \\
\Sthree    & 5   & 157 & 0.46 \\
\hline 
\end{tabular}
\caption{Parameters for the best fits to Equation~\ref{eq:hubprof} for
the density profiles plotted in Figure~\ref{fig:rhoba}.}
\label{tab:hubfit}
\end{center}
\end{table}

Figure~\ref{fig:rhoba} shows the radial gas density profiles of the
merger remnant for three mergers with impact parameters $b=0, 2.5$ and
5 (runs \Sone---\Sthree) ten sound crossing times after the merger,
when the gas has returned to hydrostatic equilibrium and the profile
has stopped evolving. The core density is normalized to $\rho_0$, and
the initial density profile is also plotted for reference.  Best-fit
values of the core radius $r_{\rm c}$ and core density $\rho_c$ are
listed in Table~\ref{tab:hubfit}. The merging process has the effect
of increasing the core radius and decreasing the core density, with a
trend towards a lower core density for the mergers with a non-zero
impact parameter, with the core gas receiving additional rotational
support.
\begin{figure}
$$\vbox{ \psfig{file=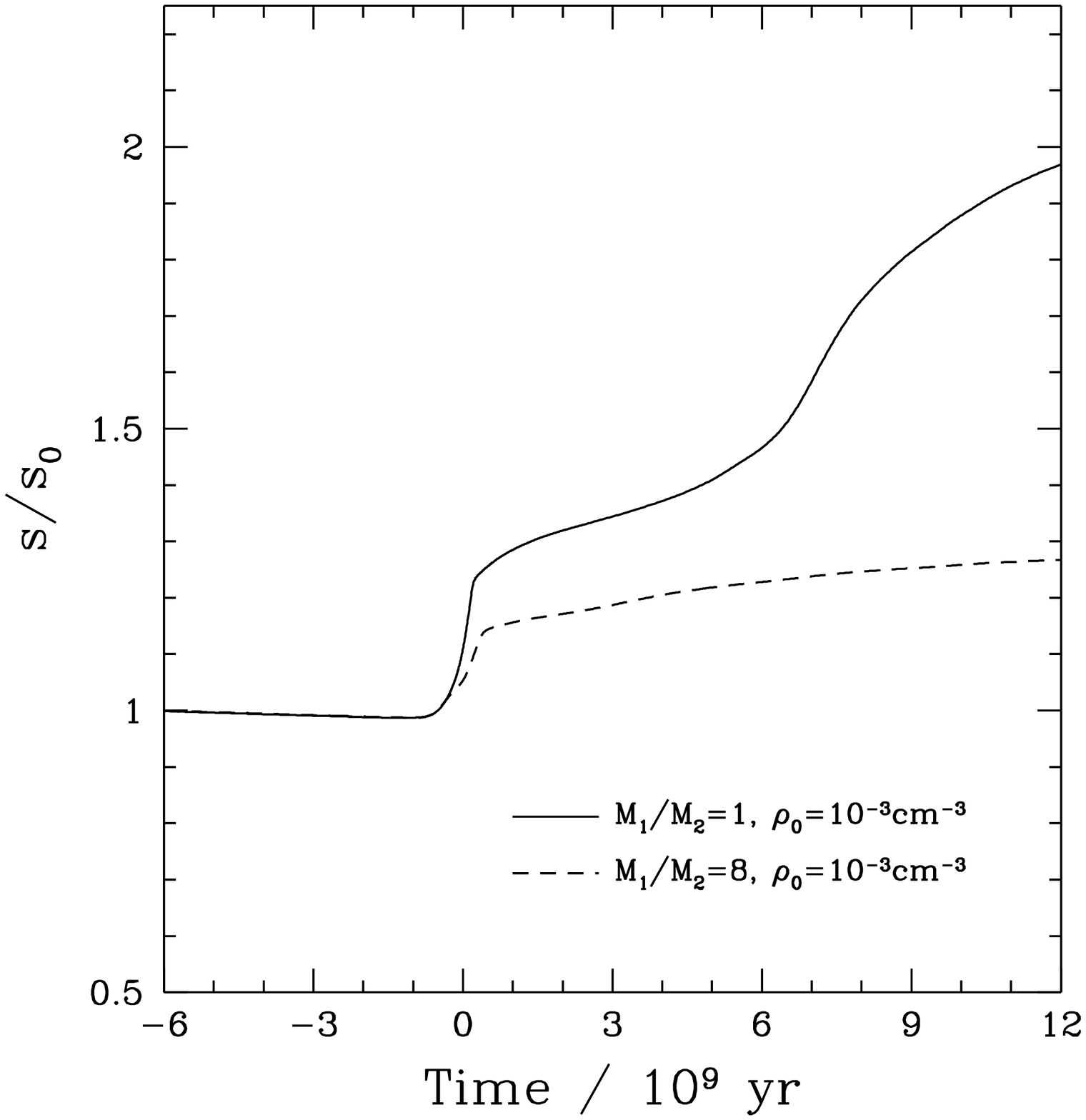,width=8.7cm} }$$
\caption{Evolution of the mean entropy of the core gas (defined in the
text) during two head-on mergers, one between two equal mass clusters
(run \Sone, solid line) and the other between two clusters with masses
in the ratio $M_1/M_2=8$ (run \Uone, dashed line). The entropy has been
normalised relative to the mean entropy at $t=-6$Gyr.}
\label{fig:sba}
\end{figure} 
The enlargement of the core is a result of an increase in the entropy
of the core gas during the merger, which can be seen in
Figure~\ref{fig:sba}. Here, we plot the mean value of the
entropy-related parameter
\begin{equation}
s_i = \epsilon_i / \rho_i^{2/3}
\end{equation}
of the 250 particles initially closest to the the core of one of the
clusters during two head-on mergers, one in which the clusters have
equal masses (run \Sone) and one in which the second cluster has one
eighth of the mass of the first (run \Uone). The cluster cores start
to interact at $t=0$, and the entropy is normalised so that it is
equal to 1 at \mbox{$t=-6$Gyr}. Initially the entropy decreases slowly
as a result of radiative cooling but jumps sharply at $t=0$ due to the
strong shock formed as the cores collide. The equal mass merger then
experiences a second, larger, jump in entropy at \mbox{$t\sim6$Gyr} as
the dark matter cores of the subclusters turn around and recollapse
(see PTC94 for a detailed discussion of this effect, which is absent
in their gas-only simulations), while the unequal mass merger
experiences only one shock. The increase in core entropy at late times
is due to the dissipation of oscillations as the core settles down
after the collision.

\subsection{X-ray Luminosity, Temperature, and the $L_{\rm x} - T_{\rm x}$ Relationship}
\label{sec:lxtx}

X-ray observatories allow precise measurement of the bolometric X-ray
luminosities ($L_{\rm bol}$) and emission-weighted temperatures
($T_{\rm ew}$) of clusters of galaxies, which are observed to be
correlated with approximately $L_{\rm bol} \propto T_{\rm ew}^3$ (Edge
\& Stewart 1991; David \etal 1993). This relationship is in conflict
with theoretical models that assume clusters form through a
self-similar gravitational collapse, which predict $M \propto T_{\rm
x}^{3/2}$ (Horner, Mushotzky \& Scharf 1999) and $L_{\rm bol} \propto
T_{\rm x}^2(1+z_f)^{3/2}$ (Scharf \& Mushotzky 1997) where $z_f$ is
the redshift at which the cluster forms, indicating that the ICM does
not share the self-similarity seen in dark-matter only models. More
recent work shows signs of a convergence between theory and
observation, as correcting the observations for the effects of cooling
flows flattens the relationship at high temperatures (Markevitch 1998;
Allen \& Fabian 1998), while more realistic numerical simulations
including radiative cooling steepen the theoretical predictions
(Muanwong \etal 2001).  In principle, the dependence on $z_f$ means
that the observed $L_{\rm bol}-T_{\rm ew}$ relationship contains
information about the evolution of clusters. However, while the
observed relationship contains significant scatter (Allen \& Fabian
1998), some of this is due to physical processes in the ICM such as
radiative cooling (Pearce \etal 2000; Muanwong \etal 2001) and
possibly preheating (\eg Lloyd-Davies, Ponman \& Cannon 2000) which
are not yet well understood. In addition merger shocks will also alter
the observed luminosity and temperature of clusters, and we examine
these effects here. Throughout this section, we calculate the
bolometric luminosity of the merging cluster as
\begin{equation}
L_{\rm bol} = \sum_i \frac{m_i \rho_i}{(\mu m_{\rm H})^2} \Lambda(T_i, Z),
\label{eq:lxbol}
\end{equation}
where the subscript $i$ denotes the sum over all gas particles, which
have temperatures $T_i$, densities $\rho_i$ and masses $m_i$. We
assume a mean molecular mass $\mu m_{\rm H} = 10^{-24}$g, and an
emissivity $\Lambda(T_i,Z)$ tabulated by Sutherland \& Dopita (1993)
with a metallicity $Z = 0.3Z_{\odot}$, where $Z_{\odot}$ is the solar
value.  The emission-weighted temperature is calculated as
\begin{equation}
T_{\rm ew} = \frac{\sum_i m_i \rho_i \Lambda(T_i,Z) T_i}{\sum_i m_i \rho_i \Lambda(T_i,Z)}.
\label{eq:tew}
\end{equation}

\begin{figure}
$$\vbox{ \psfig{file=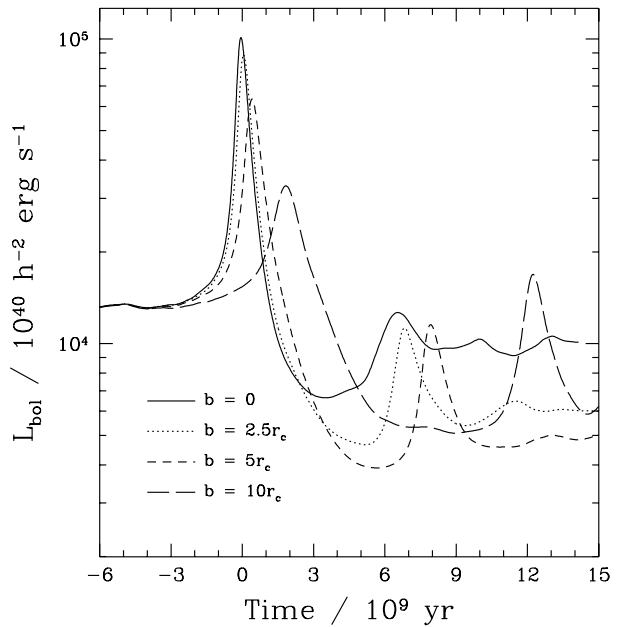,width=8.7cm} }$$
\caption{Evolution of the bolometric X-ray luminosity, defined by
Equation~\ref{eq:lxbol}, during the merger of two equal mass
systems. Four simulations are shown, with impact parameters $b=0$ (run
\Sone, solid line), 2.5 (run \Stwo, dotted line), 5 (run \Sthree,
short-dashed line) and 10. (run \Sfour, long-dashed line). Times have
been set so that $t=0$ corresponds to the peak of the bolometric X-ray
luminosity in the head-on merger.}
\label{fig:lxequal}
\end{figure} 

Figure~\ref{fig:lxequal} plots the evolution of the bolometric X-ray
luminosity during simulations \Sone---\Sfour, which follow the merger
of two equal-mass clusters at a range of impact parameters. Times are
scaled so that $t=0$ corresponds to the maximum brightening during the
head-on merger. The peak luminosity occurs when the cores of the two
clusters interact, and, in the case of head-on and nearly head-on
impacts, a large increase in luminosity is apparent, with $L_{\rm
bol}$ increasing to more than 5 times its precollision value over a
period of roughly one sound crossing time. Prior to this time no
increase in the total X-ray luminosity is apparent, despite there
being clear evidence for a merger being in progress in temperature
maps of the cluster (\eg Figure~\ref{fig:emho}, panel 3). After $t=0$,
the core of the cluster undergoes a period of expansion driven by the
dark matter, leading to a decrease in the core density and a large
($\sim 10\times$) decrease in X-ray luminosity, before experiencing a
small secondary bounce (which occurs at $t \sim 3$ for the head-on
merger, and slightly later for the simulations with a non-zero impact
parameter) and settling into hydrostatic equilibrium. The final X-ray
luminosity is lower than the sum of the emission from the two
sub-clusters due to the increase in the entropy of the core gas during
the merger. This conflicts with the results of Ricker (1998), who find
an increase in the final luminosity after the merger, but this is
likely to be an artifact of the lack of dark matter in those
simulations, and results from Ricker \& Sarazin (2001) appear to be
close to those presented here. The total X-ray luminosity of the
merger remnant is greatest in the head-on merger, with a trend towards
a lower final luminosity with increasing impact parameter in keeping
with the increase in core entropy with impact parameter seen in
Section~\ref{sec:struct}.
\begin{figure}
$$\vbox{ \psfig{file=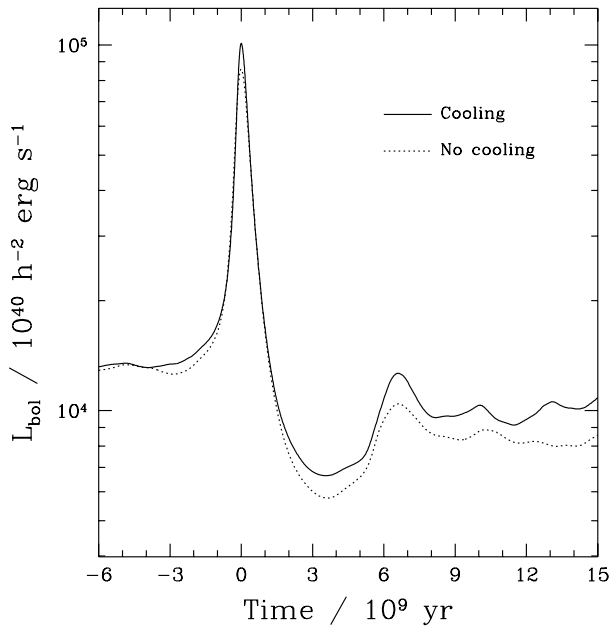,width=8.7cm} }$$
\caption{The evolution of the bolometric X-ray luminosity during two
head-on mergers of equal-mass systems, carried out both with radiative
cooling turned on (solid line) and with it turned off (dotted
line). Times are scaled as in Figure~\ref{fig:lxequal}.}
\label{fig:lxcvnc}
\end{figure} 

The effect of including radiative cooling can be seen in
Figure~\ref{fig:lxcvnc}, which plots the evolution of the bolometric
X-ray luminosity in run \Sone~against an identical merger in which
radiative cooling is not included. Radiative cooling increases the
peak bolometric luminosity during the merger by roughly 20\%, and
leads to the gas in the core of the post-merger cluster having
slightly lower entropy and consequently being more luminous than the
gas in the simulation without radiative cooling.
\begin{figure}
$$\vbox{
   \psfig{file=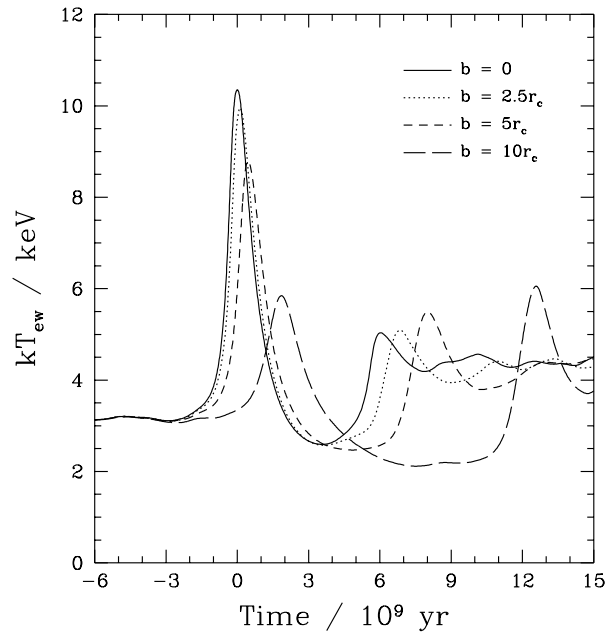,width=8.7cm}
}$$
   \caption{Evolution of the X-ray emission-weighted temperature, defined by Equation~\ref{eq:tew}, during the merger of two equal mass systems. Four simulations are shown, with impact parameters $b=0$ (run \Sone, solid line), 2.5 (run \Stwo, dotted line), 5 (run \Sthree, short-dashed line) and 10. (run \Sfour, long-dashed line). Times are scaled as in Figure~\ref{fig:lxequal}.}
\label{fig:txequal}
\end{figure} 

Figure~\ref{fig:txequal} plots the emission-weighted temperature of
the cluster during the same set of simulations. Like the X-ray
luminosity, the emission-weighted temperature increases significantly
during the merger with the effect once again being particularly marked
for mergers with a small impact parameter, for which $T_{\rm ew}$
increases by a factor of approximately 3.5. Adiabatic cooling during
the core bounce is clearly visible, with the cluster then settling
down at a higher temperature as would be expected given the higher
virial mass of the remnant.

\begin{figure}
$$\vbox{ \psfig{file=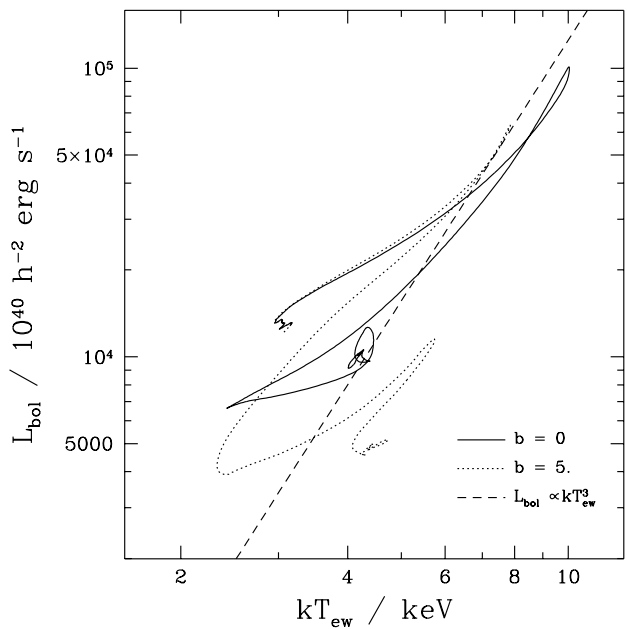,width=8.7cm} }$$
\caption{The evolution of the merging system in the \TxLx plane during
two equal-mass mergers, one head-on (run \Sone, solid line) and the
other with $b=5$ (run \Sthree, dotted line). The dashed line represents a
power-law fit to the simulations of Muanwong \etal (2001) with $L_{\rm
bol} \propto kT^3_{\rm ew}$. The high initial luminosity is a result
of plotting the sum of the X-ray emission from the two subclusters.} 
\label{fig:lxtx}
\end{figure} 

The large changes in the X-ray luminosity and emission-weighted
temperature of the cluster during the merger has implications for the
\LxTx and $M-T_{\rm ew}$ relationships.  Figure~\ref{fig:lxtx} plots
the evolution of the cluster in the \LxTx plane during two equal mass
mergers, one head-on and the other off-centre with $b=5$. The dashed
line in Figure~\ref{fig:lxtx} represents a power-law fit to the \LxTx
relationship in the simulations of Muanwong \etal (2001). The net
movement on the \TxLx plane is not that great, given that the initial
luminosity plotted in Figure~\ref{fig:lxtx} is twice the value for an
individual subcluster. The cluster becomes hotter due to the increased
virial mass, but the injection of entropy into the core limits any
increase in the luminosity of the system. However there are large
movements when the subcluster cores merge, with the system initially
following a track with $L \propto T^2$ as the core gas is compressed
adiabatically with the track steepening as the gas shocks at $t=0$,
and during this time the cluster appears to be much hotter and more
luminous than its pre-merger state. The cluster subsequently returns
to a constant entropy track as the core expands after the merger, with
both the luminosity and temperature dropping below their pre-merger
values before the core recollapses.  The implications for the
$M-T_{\rm ew}$ relation are also significant. The simulations of
Muanwong \etal (2001) find that $M \propto (kT)^{1.7}$, and so a mass
determination for the cluster based upon this relationship would vary
by a factor of $\sim 9$ depending on when the observation was
taken. The strong variations in temperature and luminosity are
relatively short lived, taking place on a timescale of around a sound
crossing time, but for large clusters this can still represent $10^9$
years or more.

\begin{figure}
$$\vbox{ \psfig{file=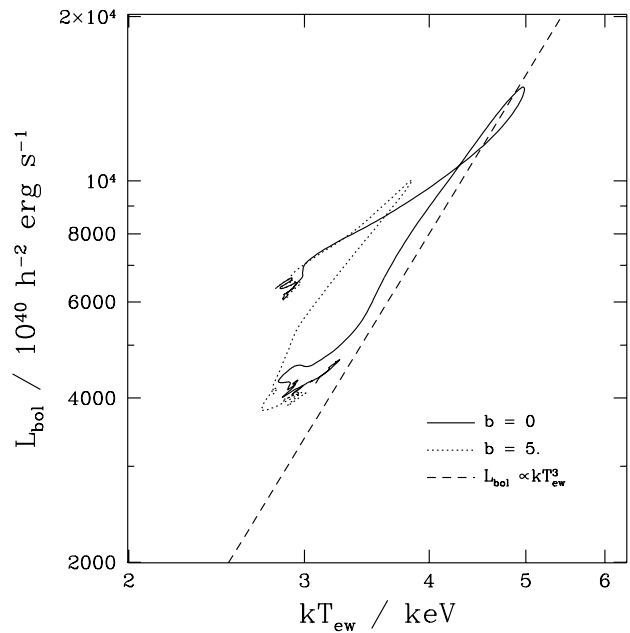,width=8.7cm} }$$
\caption{The evolution of the merging system in the \TxLx plane during
two mergers between clusters with masses in the ratio $M_1/M_2=8$ ,
one head-on (run \Uone, solid line) and the other with $b=5$ (run \Uthree,
dotted line). The dashed line represents a power-law fit to the
simulations of Muanwong \etal (2001) with $L_{\rm bol} \propto
kT^3_{\rm ew}$.}
\label{fig:lxtxue}
\end{figure} 

While equal-mass mergers will cause large fluctuations in the X-ray
luminosity and the emission-weighted temperature of a cluster of
galaxies, they are rare events. Unequal mass mergers will be much more
common, and the increase in $L_{\rm bol}$ and $T_{\rm ew}$ during
these events may prove to be a significant source of scatter in the
$L_{\rm bol}-T_{\rm ew}$ relationship. Figure~\ref{fig:lxtxue} plots
the evolution of the cluster in the $T_{\rm ew}-L_{\rm bol}$ plane
during the merger of two systems with masses in the ratio 8:1. The
maximum increase in luminosity is smaller than that seen in the
equal-mass mergers, as would be expected, but can still double during
a head-on event. The lower post-merger luminosity indicates that there
is still an injection of entropy into the core of the larger cluster,
while the emission-weighted temperature of the cluster increases by as
much as 2keV during the head-on merger.

\subsection{Cooling Flows}
\label{sec:cflow}

The simulations in Section~\ref{sec:lxtx} are concerned with clusters
with relatively low central densities ($\rho_0 = 10^{-3}$cm$^{-3}$)
and correspondingly long cooling times.  However in $\sim \!
70\%-90\%$ of observed clusters the central gas has a radiative
cooling time less than a Hubble time (Edge, Stewart \& Fabian 1992;
White, Jones \& Forman 1997). This short cooling time leads to a slow
inflow of gas to maintain pressure support known as cooling flow
(Fabian 1994), in which as much as 1000 ${\rm M}_{\odot}$yr$^{-1}$ can cool
out of the ICM (Allen \etal 1996).  There is a significant
anticorrelation between substructure in clusters and the presence of
cooling flows (Buote \& Tsai 1996) which are almost never associated
with very irregular clusters (Edge \etal 1992), but the widespread
nature of cooling flows suggests that they cannot be easily disrupted
by minor mergers, which occur relatively frequently.

\begin{figure}
$$\vbox{ \psfig{file=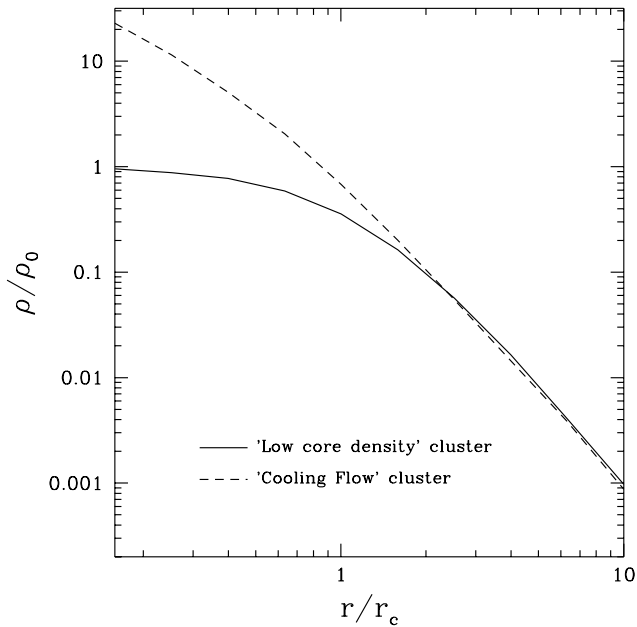,width=8.7cm} }$$
\caption{Radial density profile for our `cooling flow' cluster (dashed
line), with the initial density profile plotted for comparison
(solid line). Both profiles have been normalised to $\rho_0$, the
initial core density in the low-density cluster, and $r_c$, the core
radius in the low-density cluster.}
\label{fig:cflow}
\end{figure}

The short cooling times and high X-ray luminosities in the cores of
cooling flow clusters imply high gas densities. To assess the effects
of this, we have carried out a second series of simulations of mergers
of clusters containing cooling flows.  These simulations are set up in
a similar way to that described in Section~\ref{sec:ic}, but with the
core radius $r_c$ reduced to 40kpc to give a higher core density. The
outer density cutoff remains 1.6Mpc from the centre of the cluster,
and so now $R=40r_c$. This density profile is no longer static as the
cooling time is now comparable to the timescale of our simulations,
and so we allow the clusters to evolve in isolation until an
approximately steady state is reached (\ie the density profile in the
core has stopped evolving) before bringing them together prior to
merging. The density profile of these `cooling flow' clusters after
this initial period of relaxation is shown in Figure~\ref{fig:cflow},
with the core density being approximately 0.02cm$^{-3}$ and the
central cooling time being around $2.5\times10^9$ years. Prior to the
merger the cooling flow is depositing mass at a steady rate of just
over $80 {\rm M}_{\odot} {\rm yr}^{-1}$.

\begin{figure}
$$\vbox{ \psfig{file=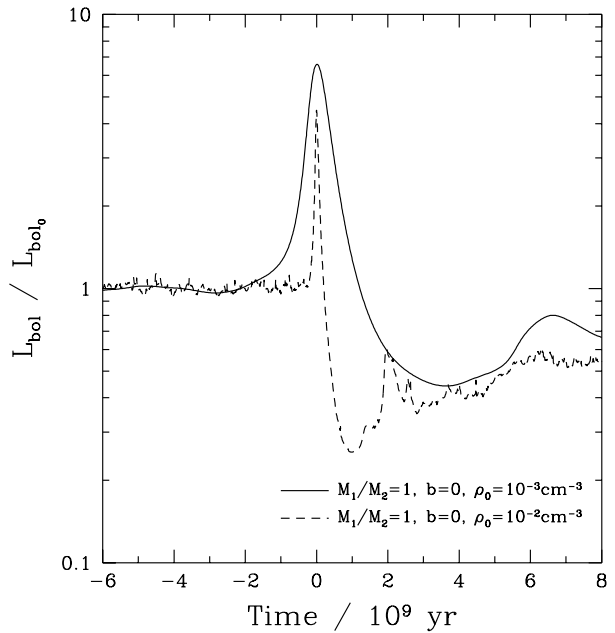,width=8.7cm} }$$
\caption{Evolution of the bolometric X-ray luminosity, defined by
Equation~\ref{eq:lxbol}, during the head-on merger of two equal mass
systems. The solid line represents the merger of two systems with core
densities of $10^{-3}$cm$^{-3}$ (run \Sone), while the dashed line
represents the merger of two `cooling flow' clusters with central
densities of $10^{-2}$cm$^{-3}$ (run \Cone). The luminosities have been
normalized to their values at $t=-6$Gyr for comparison, as the
high density of the gas in the core of the cooling flow cluster makes
it much more luminous than the clusters examined in
Section~\ref{sec:lxtx}. Times are scaled as in
Figure~\ref{fig:lxequal}.}
\label{fig:lxcf}
\end{figure} 

The evolution of the bolometric X-ray luminosity with time, shown in
Figure~\ref{fig:lxcf}, is noticeably different from the low core
density mergers examined in the previous section.  The two X-ray
brightness curves have been normalized to their luminosity at
\mbox{$t=-6$Gyr} for comparison, as the cooling-flow cluster otherwise
has a much higher luminosity due to its higher core density. Unlike
the low core density cluster, the cooling flow cluster brightens only
briefly, although the peak luminosity is only slightly lower than in
the low density simulation. In addition, the second core bounce
happens much more rapidly, and the cluster dims by a factor of more
than 30 between bounces. These effects are a result of the different
density profiles of the two clusters. The X-ray luminosity of the two
clusters is dominated by emission from the densest core gas, and will
increase when that gas is compressed. The strongly-peaked density
profile of the cooling-flow cluster will therefore increase its
luminosity when the very centre of the two clusters interact, whereas
the flat, constant density core of the low-density cluster will
experience a much more prolonged increase in luminosity. In addition,
the more centrally concentrated mass distribution in the cooling flow
cluster will relax more efficiently, leading to a shorter time between
core bounces (the same effect can be seen in Pearce \etal 1993, who
examine mergers between collisionless systems having Hubble density
profiles with $s$=2, 3 and 4). The path of the cooling flow cluster in
the \TxLx plane is similar to that seen in the previous section,
although there is continuing evolution at late times as the dense core
gas radiates away the energy gained in the merger.

\begin{figure}
$$\vbox{ \psfig{file=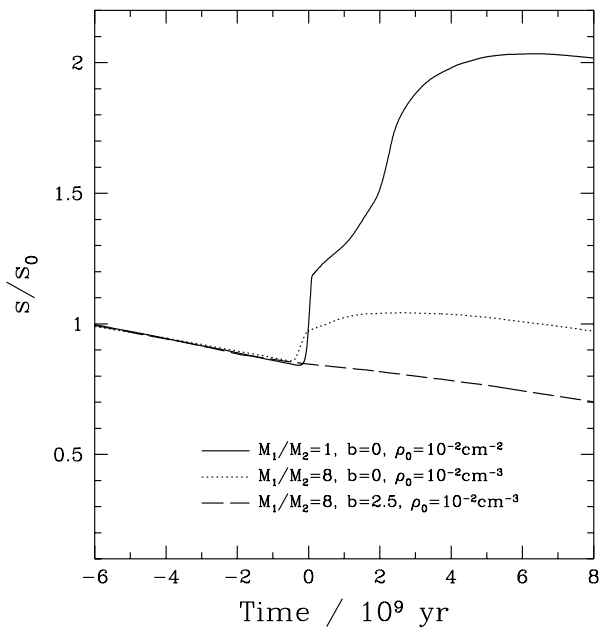,width=8.7cm} }$$ \caption{The
evolution of the mean entropy of the core gas (defined as in
Figure~\ref{fig:sba}) during three mergers between `cooling flow'
clusters, including a head-on merger between equal-mass systems (run
\Cone, solid line) and two mergers between two clusters with masses in
the ratio $M_1/M_2=8$, firstly head-on (run \Ctwo, dotted line)
and then with an impact parameter $b=2.5$ (run \Cthree, dashed
line). Each run has been normalised relative to the mean entropy at
$t=-6$Gyr.}
\label{fig:svst}
\end{figure} 

The unequal-mass merger in which the major cluster has a high density
core show little evolution in the \TxLx plane. The head-on merger (run
\Ctwo) shows a very brief increase in luminosity of roughly a factor
of two and an increase in the emission-weighted temperature of around
1.5keV, while an off-centre merger with $b=2.5$ (run \Cthree) shows
only a small ($\sim 10\%$) increase in luminosity and a
barely-noticeable increase in the emission-weighted temperature ($\sim
0.2$keV). The implication is that these unequal-mass mergers have only
a small effect on the core gas of cooling flow clusters, especially when
the merger is off-centre, and this can be clearly seen in
Figure~\ref{fig:svst}, in which the evolution of the entropy of the
gas in the core of the cooling flow cluster is plotted. In the head-on
equal-mass merger the core gas receives a large increase in entropy,
suggesting that the whole core has been disrupted, and the cooling
time in the core jumps to around $10^{10}$ years. The actual mass
deposition rate, measured from the number of gas particles cooling to
$10^4$K each timestep, drops to zero as the cores of the two clusters
collide, increasing slowly after the merger, although even after 10Gyr
the mass deposition rate is still less than $20 {\rm M}_{\odot} {\rm
yr}^{-1}$.  However, the apparent mass deposition rate derived from
the luminosity and temperature within the cooling radius
\begin{equation}
\dot{M} = \frac{2\mu m_{\rm H}}{5k_{\rm B}T}L_{r<r_{\rm cool}}
\end{equation}
(Fabian 1994) increases briefly during the merger and does not provide
an accurate estimate of the mass cooling out of the ICM until the
cluster has returned to hydrostatic equilibrium.

The head-on unequal-mass merger also shows a jump in entropy, and
although the cooling time of the core gas doubles to around
$4\times10^9$ years the energy injected into the core rapidly starts
to be radiated away. The mass deposition rate again drops sharply
during the merger, but recovers to around $55 {\rm M}_{\odot} {\rm
yr}^{-1}$ within 4--5 Gyr. In contrast to the head-on mergers, the
unequal-mass off-centre merger does not display any signs of the core
gas being shocked, indicating that the bow-shock around the infalling
subclump cannot have penetrated all the way into the core of the
cluster. However the entropy of the core gas is decreasing more slowly
after the merger, reflecting a slight increase in the cooling time to
around $3\times10^9$ years accompanied by a drop of around $30\%$ in
the mass deposition rate.  The increase in the central cooling time is
a result of ram pressure from the infalling subcluster displacing gas
from the core of the cluster (Fabian \& Daines 1991; Gomez \etal
2000). Ram pressure also accounts for the bulk of the disruption to
the head-on, unequal-mass merger, as the increase in core entropy
alone is insufficient to account for the doubling of the cooling time
in the cluster core.

The robustness of high-density cores to the effects of what should be
a relatively common merger event is of interest to the survival of
cooling flows in clusters of galaxies.  Our simulations suggest that a
major merger between clusters of equal mass will completely disrupt
any cooling flow activity, but cooling flows, once established, may
prove resilient to mergers with lower-mass subclumps, especially if
they are off-centre.  While the cooling time in the low core density
simulations remains greater than a Hubble time during almost all of
the merger process, dipping below $10^{10}$ years only during the
maximum compression of the cores, the cooling time in the high-core
density clusters remains below a Hubble time throughout the
process. In the case of the equal-mass merger the increase in core
entropy and cooling time is large, and is enough to both disrupt the
cooling flow and to prevent it from restarting soon after the merger,
while the increase in the cooling times in the unequal-mass mergers are
much less, and would indicate that the cooling flows would either
experience minimal disruption or would rapidly be reestablished.

\subsection{Mixing}
\label{sec:mixing}

Observations of the abundance of metals in the ICM suggest that
negative radial metallicity gradients are common in clusters of
galaxies (Irwin \& Bregman 2001 and references therein). Significant
gradients are most common in clusters containing cooling flows, while
clusters with no cooling flow also tend to have a flat metallicity
gradient, although this is not universal. The metallicity gradient is
thought to be a result of either galactic winds or the stripping of
metal-enriched gas from galaxies, and Irwin \& Bregman (2001) suggest
that the non-cooling-flow clusters have experienced a merger that both
disrupted the cooling flow and mixed the ICM, thus erasing the
metallicity gradient.

We examine the degree of mixing in the merger remnant in our
simulations by smoothing the gas particle distribution on to a grid,
and measuring a quantity ${\cal M}$ in each cell, where
\begin{equation}
{\cal M} = 1 - \left | \frac{n_1 - n_2}{n_1 + n_2} \right |
\label{eq:mix}
\end{equation}
and $n_1$ and $n_2$ are the number of particles in the cell which were
in cluster 1 and 2 respectively at the start of the simulation.  The
cell size is set so that each cell will, on average, contain at least
ten particles, with the cells in the core containing more than this
number, although this is not always true in the low-density outer
regions of the cluster which are represented by relatively few
particles. In a well-mixed cell $n_1$ and $n_2$ will be equal and so
${\cal M} \rightarrow 1$, while in a cell in which no mixing has
occurred ${\cal M} \rightarrow 0$.

\begin{figure}
$$\vbox{ \psfig{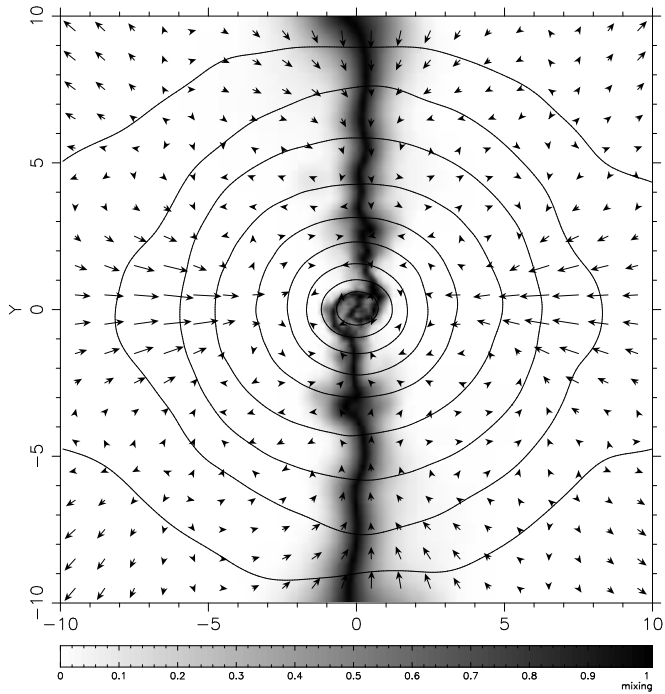} }$$
\caption{Mixing 5Gyr after a head-on merger between two
equal mass clusters (run \Sone). Grayscales show the degree of mixing (defined in
the text), contours represent logarithmic only X-ray isophotes and arrows
represent the velocity field.}
\label{fig:emhomix}
\end{figure} 
\begin{figure}
$$\vbox{ \psfig{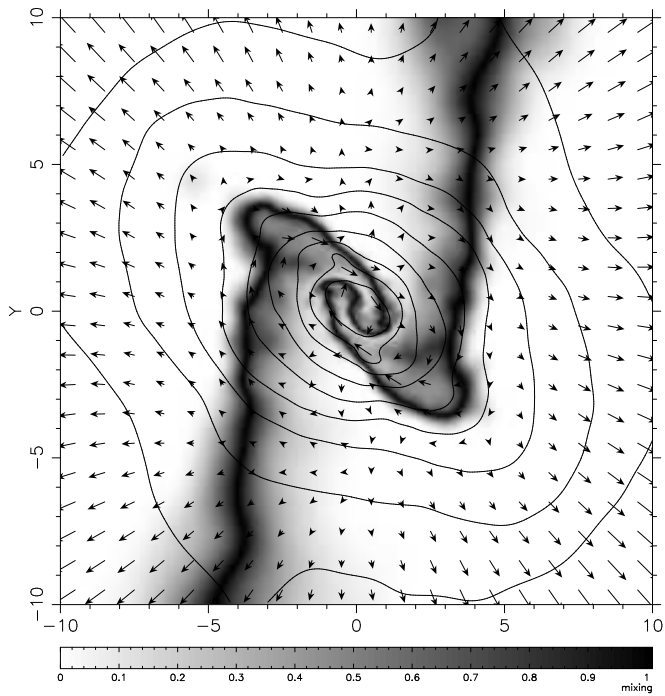} }$$
\caption{Mixing 5Gyr after a off-centre merger with
impact parameter $b=5$ between two equal mass clusters (run
\Sthree). Grayscales, contours and arrows are as in Figure~\ref{fig:emhomix}.}
\label{fig:emb5mix}
\end{figure} 

The degree of mixing 5Gyr after a head-on, equal-mass merger is
shown in Figure~\ref{fig:emhomix}. Very little mixing of the ICM has
taken place, with the mixing in the centre of the cluster only appearing
at around $t=4$Gyr, apparently driven by the infall of material back
along the merger axis rather than by the merger itself. Much more
mixing is visible in Figure~\ref{fig:emb5mix}, which shows the state
of an off-centre merger with $b=5$. This is also plotted at $t=5$Gyr,
although in this case the cluster is still settling back to a steady
state, as can be seen in the distortion of the X-ray isophotes.  The
angular momentum in the system means that the core of the merger
remnant is well mixed, although there is still little mixing of the
ICM at $r>b$.

\begin{figure}
$$\vbox{ \psfig{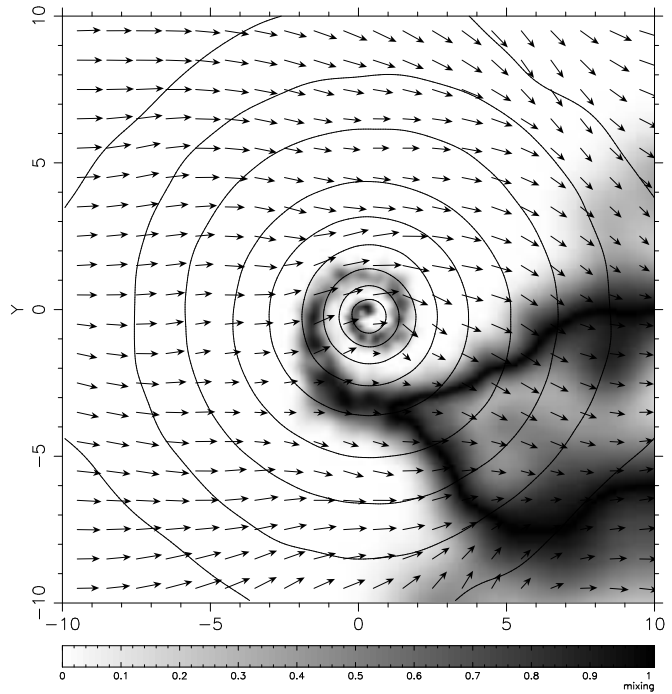} }$$
\caption{Mixing 5Gyr after a off-centre merger with
impact parameter $b=5$ between two clusters with masses in the ration
$M_1/M_2=8$ (run \Uthree). Grayscales, contours and arrows are as in
Figure~\ref{fig:emhomix}.}
\label{fig:m8b5mix}
\end{figure} 

Finally, Figure~\ref{fig:m8b5mix} shows the degree of mixing during an
off-centre merger in which the main cluster has a mass eight times
that of the infalling subcluster. In this case, low-density gas is
rapidly stripped from the subcluster and mixing occurs throughout the
outer layers of the main cluster. However, the subcluster core is less
efficiently stripped, and can be seen spiraling all the way into the
core of the main cluster. There is a clear difference between the
final states of the equal mass and unequal mass mergers. Very little
mixing occured in the outer layers of the cluster during the equal
mass mergers, although the off-centre mergers mixed the cores
efficiently, while the unequal mass merger caused very little mixing
in the core but lead to significant mixing in the outer regions of the
cluster.

While none of these mergers mixed the ICM globally, our simulations
suggest that mergers of the magnitude necessary to disrupt a cooling
flow can efficiently mix the cores of clusters on the scales
accessible to X-ray observations (in general $<50\%$ of the virial
radius) if the merger is off-centre. In contrast, the unequal mass
merger was unable to mix the core gas efficiently, but may cause
locally increased abundances such as that seen by Arnaud \etal (1994)
in the Perseus cluster and are consistent with the presence of a
cooling flow, which will not necessarily be disrupted during a minor
merger.

\subsection{Numerical Resolution}
\label{sec:res}

\begin{table}
\begin{center}
\begin{tabular}{|c|c|c|c|}
\hline
Run no.& $N_{\rm tot}$ & $m_{\rm dm} ({\rm M}_{\odot})$& $m_{\rm gas}
({\rm M}_{\odot})$\\

\hline
\Sone   & 131072 & $1.25 \times 10^{10}$ & $1.13\times10^9$ \\
\Rone   & 32768  & $5\times 10^{10}$   & $4.5\times10^9$  \\
\Rtwo   & 8192   & $2\times 10^{11}$  & $1.8\times 10^{10}$  \\
\hline 
\end{tabular}
\caption{Particle numbers and masses for the simulations discussed in
Section~\ref{sec:res}.}
\label{tab:resol}
\end{center}
\end{table}

Quantifying the effects of limited resolution is always important in
numerical simulations. To assess the impact that resolution has on the
results presented here, we have carried out two further simulations of
head-on mergers between equal mass systems in which the total number
of particles have been reduced by a factor of 4 (run \Rone) and 16 (run
\Rtwo), giving a total of 32768 and 8192 particles respectively, with
the cluster properties kept constant by increasing the mass of the
remaining particles. Details of the simulation parameters are shown in
Table~\ref{tab:resol}.

\begin{figure}
$$\vbox{ \psfig{file=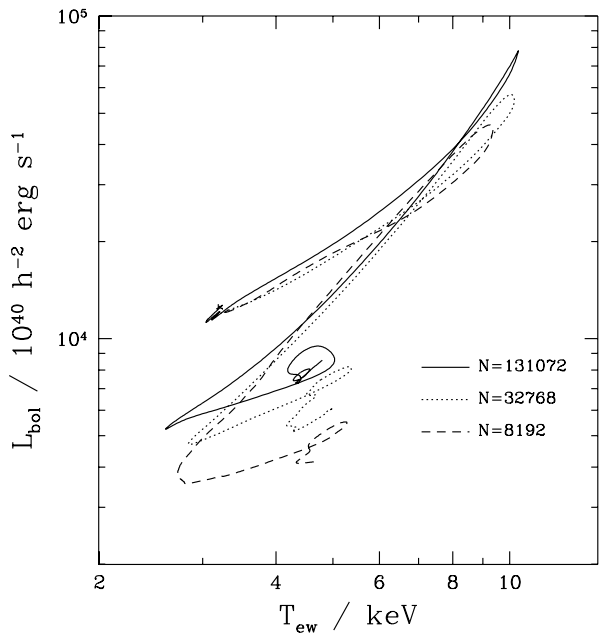,width=8.7cm} }$$
\caption{Evolution of the merging system on the \TxLx plane during 
during three head-on mergers between equal mass systems. The solid
line represents a simulation in which each cluster has 32768 particles
each of gas and dark matter (run \Sone), the dotted line a simulation
with 8192 particles of each type (run \Rone) and the dashed line a
simulation with 2048 particles of each type (run \Rtwo). The total mass
of the systems is kept constant by increasing the particle masses in
the lower resolution runs. }
\label{fig:convlxtx}
\end{figure} 

The evolution of the clusters in the \TxLx plane during these
simulations is shown in Figure~\ref{fig:convlxtx}. The convergence
between runs is good, with the evolution of the system being similar
in all three simulations. The main effect of lowering the resolution
is a systematic decrease in the luminosity of the merger remnant. In
our low-resolution runs less energy is deposited in the halo gas than
in the full runs and the core gas is shocked to a greater extent,
leading to a broader, higher-entropy core in the final cluster, a
result also seen in the resolution testing carried out in
PTC94. The differences in the emission-weighted temperature are small.

If the particle number is reduced further then the agreement between
simulations worsens, and we would agree with the conclusions of PTC94
that 1000 gas particles per cluster represent a minimum for modelling
the bulk properties of a merging system. However, more than $10^4$ gas
particles per cluster are required to produce well-resolved
temperature and X-ray surface brightness maps such as those shown in
Figures~\ref{fig:emho} and~\ref{fig:emb5}, in agreement with Steinmetz
\& M\"{u}ller (1993).

We conclude that the effects of resolution are of limited significance
in our simulations. The evolution of the cluster in the \TxLx plane
has converged to a consistent result, differing only in detail as the
resolution is increased. Equally, we have sufficient resolution to
follow the propagation of shocks in the ICM acceptably, although here
our resolution is closer to the minimum required for the problem.

\section{Discussion}
\label{sec:discuss}

It is clear from the results presented here that mergers can alter the
global properties of clusters of galaxies. During a merger,
hydrodynamic shocks dissipate much of the kinetic energy of the impact
into the ICM, leading to a departure from hydrostatic equilibrium that
can last for several Gyr. Mergers lead to many potentially observable
effects, including strong temperature gradients across shocks,
distortion in the X-ray isophotes, multiple peaks in the X-ray surface
brightness and high-velocity bulk flows in the ICM. Despite this,
identifying merging clusters is not necessarily easy, as both
projection effects and the limits of X-ray observations can serve to
mask many of the signs of an ongoing merger. While X-ray substructure
is clearly visible both before and after a merger when viewed
perpendicular to the merger axis (as can be seen in
Figures~\ref{fig:emho} and~\ref{fig:emb5}), the X-ray isophotes appear
spherically symmetric throughout the merger when viewed along the
merger axis. Equally, the shocks so apparent in temperature maps when
viewed perpendicular to the merger are not obvious when viewed along
the merger axis, with the cluster just appearing to have a steep
temperature gradient as the cores of the two subclusters merge. While
this departure from isothermality is itself a good indicator of a
merger, it requires that the temperature profile is both sufficiently
well resolved for it to be apparent and is not unduly affected by the
effects of the merger, as the X-ray deprojection method (\eg Arnaud
1988) assumes that the cluster is spherically-symmetric and in
hydrostatic equilibrium. Temperature maps, derived directly from the
X-ray hardness ratio, will therefore be a better indicator of an
ongoing merger than deprojected temperature profiles. If the cluster
is poorly resolved, it will simply appear to be well relaxed, hot and
luminous, although evidence for a merger will still be seen in the
distribution of the peculiar velocities of the galaxies in the two
clusters, which can be expected to be significantly bimodal during a
merger. At intermediate angles, projection effects will serve to
smooth distortions in the X-ray isophotes and reduce the visibility of
the merger in the X-ray waveband, although the effects are still
apparent. 
%While we do not examine mergers with higher impact
%velocities, PTC94 find that a higher merger velocity strengthens the
%merger shock resulting in a greater compression of the core and a
%higher peak luminosity. In addition, the entropy of the core gas will
%be higher and therefore the cluster will have a broader, lower density
%core.

The merging process has major implications for statistical studies of
clusters. As we have shown in Section~\ref{sec:lxtx}, even relatively
minor ($M_1/M_2=8$) mergers can introduce a significant scatter in the
observed \LxTx relationship, while estimates of the mass of clusters
based on the $M-T_{\rm ew}$ relationship (Horner \etal 1999) will
suffer large uncertainties during and immediately after a merger.
Mass estimates based on assumptions of hydrostatic equilibrium (\eg
Fabricant, Lecar \& Gorenstein 1980) will also be strongly affected as
the departures from equilibrium can last for several Gyr after the
merger, during which time many of the readily-observed signs of the
merger have faded, and Roettiger \etal (1996) estimate that errors in
the mass estimate may reach 50\% in the 2Gyr following a merger. Edge
\etal (1992) suggest that clusters will typically undergo a merger
every 2-4Gyr, similar to the timescale for our clusters to return to
hydrostatic equilibrium, implying that truly relaxed clusters may be
uncommon. The large increase in X-ray luminosity during a merger will
also introduce strong selection effects into cluster surveys, and the
clusters found at high redshift may not be a representative sample.
The increases in X-ray luminosity and temperature during a major
merger will strongly influence the statistics of the hottest and most
luminous clusters of galaxies. This has been confirmed by \Chandra
temperature maps of two of the hottest clusters known, 1E 0657-56
(Tucker \etal 1998) and A2163 (David \etal 1993; Markevitch \&
Vikhlinin 2001), both of which show strong merger shocks. In addition,
unusual clusters like A851, which is cooler and less luminous than
it's richness would suggest, may be undergoing the post-merger core
expansion (Schindler \& Wambsganss 1996). 

The situation is more complex when cooling flow clusters are
considered. We find that major mergers will almost certainly disrupt
cooling flows to the extent that they cannot restart within a Hubble
time, consistent with the lack of cooling flows in irregular clusters
(Edge \etal 1992), but our unequal mass mergers have less of an effect
on the cooling flow. A head-on merger still disturbed the cooling
flow, although it restarted within a relatively short period, while an
off-centre merger had little impact on the flow. The survival of the
cooling flows during minor mergers is supported by a growing body of
observations of cooling flow clusters that also show evidence for a
late-stage merger (\eg A2142; Markevitch \etal 2000), as well as
merging clusters with small cooling flow cores (\eg A2065; Markevitch
\etal 1999) which suggest that the cooling flow has been disturbed but
has survived, at least until the present day.

Finally, we examined the degree to which mergers can mix the
ICM. While the ICM is never well-mixed globally, our simulations
suggest that major off-centre mergers can efficiently mix the core
gas, although there was very little mixing during a head-on
merger. Equal mass mergers will also disrupt any cooling flow in the
cluster, and may explain why metallicity gradients are rare in
clusters without a cooling flow (Irwin \& Bregman 2001). The unequal
mass merger introduced a high degree of mixing in the outer layers of
the main cluster as the low-density gas in the subclump is rapidly
stripped away, but had little effect on the core gas.

A feature of unequal-mass mergers that we have not touched on in this
paper is the survival of the cores of subclusters during mergers.
Sharp surface brightness discontinuities have been observed by
\Chandra in merging clusters of galaxies such as Abell 2142
(Markevitch \etal 2000) and Abell 3667 (Vikhlinin \etal 2001), and it
has been suggested that these features, termed ``cold fronts'', are
low-entropy gas from the core of a subcluster that had so far survived
the merging process (Markevitch \etal 2000). We find that subcluster
cores can survive off-centre mergers, and result in features very
similar to those observed by \Chandra; this will be examined further
in a subsequent paper.

\section{Conclusions}
\label{sec:conc}

We have presented results from high-resolution AP$^3$M+SPH simulations
of merging clusters of galaxies. Our principle conclusions are; 

\begin{enumerate}

\item Mergers lead to many potentially observable effects, including
strong temperature gradients across shocks, distortion in the X-ray
isophotes, multiple peaks in the X-ray surface brightness and
high-velocity bulk flows in the ICM.

\item An increase in the entropy of the core gas during the merger
leads to a broader, lower-density core in the post-merger cluster. In
addition, off-centre mergers can give the core additional rotational
support.

\item The compression and shocking of the core gas during a merger can
lead to large increases in the bolometric X-ray luminosity and the
emission-weighted temperature of the cluster. This will have a
significant impact on statistical studies of galaxy clusters. 
Immediately after the cores of the subclusters interact, the
core of the cluster undergoes a period of expansion driven by the
collisionless dark matter, cooling adiabatically and decreasing in
luminosity by more than an order of magnitude, before the cores turn
around and recollapse.

\item Radiative cooling leads to an increase of $\sim 20\%$ in the
X-ray luminosity during and after a merger when compared to
simulations without cooling, even if central densities are relatively
low ($\rho_0 = 10^{-3}$cm$^{-3}$) and the cooling times in the pre-
and post-merger objects are greater than a Hubble time.

\item In a major merger the cluster can be out of hydrostatic
equilibrium for several Gyr, during which time the clear observational
signs of a merger will become less apparent. Brief departures from
hydrostatic equilibrium were also seen in simulations of unequal mass
mergers.

\item Cooling flows will be completely disrupted during major mergers,
and are unlikely to be able to restart within a Hubble time. However,
minor mergers had less of an effect on a cooling flow, which was
either unaffected by the merger or was able to restart rapidly. Ram
pressure is more effective at disturbing the cooling flow in minor
mergers than the merger shock. During a merger the mass deposition
rate inferred from observations is a poor estimate of the actual mass
cooling out of the ICM.

\item Major off-centre mergers effectively mixed the gas in clusters
within a radius roughly equal to $b$, the impact parameter.  Head on
mergers and unequal-mass mergers had little effect. None of the
mergers examined here effectively mixing the ICM globally.

\end{enumerate}

\section*{Acknowledgments}

BWR acknowledges the support of a PPARC postgraduate studentship. PAT
is a PPARC Lecturer Fellow. We thank the referee for helpful comments
that have improved this paper.

%\section*{References}

\vfill

\end{document}